\documentstyle[a4,psfrag,graphicx,aps,pre,amstex,epsfig]{revtex}

\draft

\def\bra{\langle}
\def\ket{\rangle}
\newcommand{\ch}{\chi\rule[-1ex]{0mm}{2ex}}
\DeclareMathSymbol{\MMM}{\mathbin}{AMSb}{'115}
\DeclareMathSymbol{\PPP}{\mathbin}{AMSb}{'120}
\DeclareMathSymbol{\DDD}{\mathbin}{AMSb}{'104}
\DeclareMathSymbol{\ZZZ}{\mathbin}{AMSb}{'132}

\def\FIGURE#1#2{
  \clearpage
  \voffset-2cm
  \begin{center}
    {\small\sffamily C.Manwart and R.Hilfer \hfill Figure \ref{fig:#1}}\\[3cm] 
    \epsfig{figure=./#1.ps,width={#2\linewidth}}
  \end{center}
}

\newenvironment{CAPTION}
{  \newcounter{fig}
   \begin{list}
     {\textbf{Figure \arabic{fig}:}}
     {\usecounter{fig}
       \setlength{\labelwidth}{2.2cm}
       \setlength{\labelsep}{0.3cm}
       \setlength{\itemindent}{0pt}
       \setlength{\leftmargin}{2.5cm}
       \setlength{\rightmargin}{0cm}
       \setlength{\parsep}{0.5ex plus0.2ex minus0.1ex}
       \setlength{\itemsep}{0ex plus0.2ex}
     }
}
{\end{list}}

\def\CAP#1#2{\item \label{fig:#1} {#2} }

\begin{document}
\bibliographystyle{prsty}

\title{On the Reconstruction of Random Media using Monte Carlo Methods}

\author{C. Manwar$t^1$ and R. Hilfe$r^{1,2}$} 

\address{$\ ^1$ Institut f\"ur Computeranwendungen 1, Universit\"at
  Stuttgart, 70569 Stuttgart, Germany\\
  $\ ^2$ Institut f\"ur Physik, Universit\"at Mainz, 55099 Mainz, Germany} 
\date{\today}

\maketitle

\begin{abstract}
  A simulated annealing algorithm is applied to the reconstruction of
  two-dimensional porous media with prescribed correlation functions.
  The experimental correlation function of an isotropic sample of
  Fontainebleau sandstone and a synthetic correlation function with
  damped oscillations are used in the reconstructions. To reduce the
  numerical effort we follow a proposal suggesting to evaluate the
  correlation functions only along certain directions. The results
  show, that this simplification yields significantly different
  micro-structures as compared to a full evaluation of the correlation
  function. In particular we find that the simplified reconstruction
  method introduces an artificial anisotropy that is originally not
  present.
\end{abstract}

\bigskip
\begin{tabbing}
  PACS: \= 61.43.G \hspace*{2ex}\= (Porous materials; structure),\\
  \> 61.43.j \> (Disordered solids),\\
  \> 81.05.Rm \> (Porous materials; granular materials),\\
\end{tabbing}

\clearpage
\section{Introduction}
A better understanding of the transport properties of random media,
such as fluid flow in sandstones or electrical conductivity of
composites requires the micro-structure as input
\cite{hil91d,hil92a,hil93b,hil94b,hil95d,hil98a}. Digitized,
three-dimensional micro-structures of natural sandstones are difficult
and expensive to obtain \cite{STJLJA94}. Thus there is a need for simulation
algorithms that are able to provide representative micro-structures
from statistical probability functions.

Recently various algorithms have been proposed for the reconstruction
of random micro-structures \cite{YT98a,rob97,adl92,qui84}.  In this
paper we investigate a simulated annealing method that enforces
agreement between the correlation functions of the original and the
reconstructed micro-structure \cite{YT98a}.  To save computation time
the authors of \cite{YT98a} have evaluated the correlation functions
only in certain directions assuming isotropy of the medium. The
objective of this paper is to study the effect of this simplification
on the final, reconstructed configurations.  We test the effects on
two examples: (i) the correlation function of a Fontainebleau
sandstone and (ii) an artificial correlation function with damped
oscillations. The results show, that at least for the oscillating
correlation function this yields significantly different
configurations. More importantly, the reconstructions are anisotropic
as a result of the simplified evaluation of the correlation functions.

\section{The reconstruction method}
We follow the reconstruction algorithm proposed in \cite{YT98a}.  The
reconstruction is performed on a $d$-dimensional hyper-cubic lattice
$\ZZZ^d$ ($d=2$ for the results presented below). Whether a lattice
point lies within pore space or matrix space is indicated by the
characteristic function
\begin{equation}
  \ch(\vec x)= \ch(x_1, x_2, \dots , x_d)  = \left\{
  \begin{array}[]{l}
    0 \quad \hbox{for} \quad \vec x \in \PPP \\
    1 \quad \hbox{for} \quad \vec x \in \MMM 
  \end{array}
  \right.
\end{equation}
with $x_i = 0,1,\dots,M_i-1$ where $\PPP$ denotes the pore space and
$\MMM$ the matrix space of a two phase porous medium. The $x_i$ are in
units of the lattice spacing $a$. The porosity $\phi$ is given as
$\phi = \frac{1}{N}\sum_{j=1}^N\left(1-\ch(\vec x_j)\right)$ where
$N=\prod_{i=1}^d M_i$ is the total number of lattice sites.

Simulated annealing is an iterative technique for combinatorial
optimization problems. The iteration steps are denoted by a subscript
$t$.  The optimum is found by lowering a fictitious temperature $T_t$
that controls the acceptance or rejection of configurations with
"energy" (or cost function) $E_t$. The energy function used in our
simulations is defined as
\begin{equation} \label{eq:energy}
  E_t = \sum_{k=1}^J w_k \sum_{\vec r \in \DDD_k} \left( g_t^k(\vec r) - g_{\rm
      ref}^k(\vec r) \right)^2
\end{equation}
where $\DDD_k \subset \ZZZ^d$ is a subset of lattice points and
$g_t^k$ is the $k$th function of a set of $J$ statistical probability
functions calculated for the configuration of step $t$. For example
$g_t^k$ may be a $k$-point correlation function. The real valued
factor $w_k$ is a weight for the $k$th function. Hence, the energy can
be understood as a measure for the deviations of the probability
functions $g_t^k$ from predefined reference functions $g_{\rm ref}^k$.

The simulated annealing algorithm consists of the following steps.
\begin{enumerate}
\item Initialization: The 0's and 1's are randomly distributed with
  given porosity $\phi$.
\item Two lattice points of different phase are chosen at random and
  exchanged. In this way the porosity $\phi$ is conserved.
\item The "energy" $E_t$ of the current configuration is calculated
  according to Equation (\ref{eq:energy}).
\item The "temperature" $T_t$ is adjusted according to a fixed cooling
  schedule.
\item The new configuration created by the exchange of the two points
  is accepted with probability
  \begin{equation} \label{eq:p}
    p = \min\left(1, \exp\left(- \frac{E_t - E_{t-1}}{T_t}\right) \right)
    .
  \end{equation}
  In case of rejection, the two points are restored and the old
  configuration is left unchanged. 
\item Return to step 2.
\end{enumerate}

As can be seen from Equation (\ref{eq:p}), configurations with lower
energy are immediately accepted while the acceptance of a
configuration with higher energy is controlled by the temperature $T$.
In order to obtain a configuration with minimal energy $E$ or, in
other words, a configuration with minimal deviations of the
probability functions $g^k$ from their reference functions $g_{\rm
  ref}^k$, the temperature $T$ has to be lowered in a suitable way.
                                
The algorithm is applicable to various functions $g^k$. Most authors
propose the two point correlation function \cite{YT98a,adl92,qui84}.
Assuming homogeneity and ergodicity, the two-point correlation
function can be defined for our case as
\begin{equation} \label{eq:acf_gen}
  g(\vec r) = \frac{\bra \ch (\vec x) \ch (\vec x + \vec r) \ket -
    (1-\phi)^2}{\phi - \phi^2}
  .
\end{equation}
where the average $\bra \dots \ket$ indicates a spatial average over
all lattice sites $\vec x$.  If $g_{t-1}(\vec r)$ is known a single
update step of the annealing process requires the recalculation of the
correlations of the two pixels that are exchanged in step 2 with all
other pixels.  The numerical effort to obtain $g_t(\vec r)$ is
therefore proportional to $N$ where $N$ is the total number of lattice
points.  In the reconstruction of three-dimensional porous media with
$N\approx 10^6 \dots 10^7$ this leads to unacceptably long calculation
times and therefore one has to find ways to speed up this calculation.
  
One possibility for reducing the numerical effort is to truncate $g$
at a value $r_c$ for which $g(\vec r) \approx 0$ for $r = |\vec r| 
\geq r_c$ holds.  Below we set $g(\vec r)=0$ for $|\vec r|\ge r_c$ where
$r_c$ is a parameter in the reconstruction.  For isotropic media,
where $g(\vec r) = g(r)$ with $|\vec r|=r$, it was suggested in
\cite{YT98a} to calculate $g(r)$ only in certain directions by setting
$\vec r = r \vec e_{k}$ where $\vec e_{k}$ is an arbitrary unit
vector. In the two-dimensional reconstructions presented below $\vec
e_k$ will be set to the radial unit vector in a polar coordinate
system $\vec e_k = \vec e_{\varphi_k} = \vec e_x \cos \varphi_k + \vec
e_y \sin \varphi_k$ where $\vec e_x$ and $\vec e_y$ are the unit
vectors of the Cartesian coordinate system and $\varphi_k$ is the
angle between $\vec e_x$ and $\vec e_{\varphi_k}$.  Hence, instead of
Equation (\ref{eq:acf_gen}) we use
\begin{equation} \label{eq:acf_simple}
  g^k(r) = \frac{\bra \ch (\vec x) \ch (\vec x + r\vec e_{k}) \ket -
    (1-\phi)^2}{\phi - \phi^2} 
\end{equation}
with $r = 0,1,\dots,r_c$ in the simplified reconstruction scheme.
Since (\ref{eq:acf_simple}) is a set of $J$ one-dimensional
correlation functions the numerical effort is reduced by a factor of
roughly $Jr_c/N$ as compared to (\ref{eq:acf_gen}).

The above algorithm is now used to reconstruct two-dimensional,
isotropic media with correlation function
\begin{equation} \label{eq:acf_osc}
  g_{\rm ref}(r) = \exp\left(-\frac{r}{8}\right)\cos(\omega r)
\end{equation}
with $r$ in units of the lattice spacing and $\omega = 1$. The same
function was used by the authors of \cite{YT98a} to exemplify their
algorithm. In the evaluation of the correlation functions periodic
boundary conditions are assumed. We use an exponential decrease of the
temperature
\begin{equation}
  T_t = \exp\left(-\frac{t}{16\cdot 10^5}\right)
  .
\end{equation}
The remaining parameters are the lattice size $M_1 = M_2 = 400$ and
the porosity $\phi=0.5$. They are the same as in \cite{YT98a}. The
following reconstructions have all been initialized with the same
random seed. The algorithm terminated when the configuration did not
change for 20000 subsequent update steps.

\section{Results}

In the first reconstruction, the correlation function was calculated
only in the horizontal and vertical direction, i.e. Equations
(\ref{eq:energy}) and (\ref{eq:acf_simple}) were used with $J=2$,
$\vec r_1 = r \vec e_0$ and $\vec r_2 =r \vec e_{\pi/2}$.  Both
correlation functions $g^1, g^2$ had the same weight $w_1=w_2=0.5$ and
the same reference function $g^1_{\rm ref} = g^2_{\rm ref}$ given in
Equation (\ref{eq:acf_osc}) was used. The correlation function was
truncated at $r_c=100$. Figure \ref{fig:figure_1} shows the final
configuration of the reconstruction. A similar pattern was found in
\cite{YT98a}. The pattern consists of stripes in direction of $\vec
e_{\pi/4}$ and in direction $\vec e_{-\pi/4}$.  The distance between
the stripes is determined by the cosine term. On a larger length
scale, the pattern organizes into several regions in which all lines
are parallel. The typical size of these regions is given by roughly 20
in view of Equation (\ref{eq:acf_osc}) and Figure \ref{fig:figure_2}.
Clearly, the pattern is not isotropic as it should be. The stripes are
preferentially directed along the directions $\vec e_{\pi/4}$ and
$\vec e_{-\pi/4}$.  Also, one expects that the oscillation frequency
$\omega$ of the correlation function in direction $\vec e_{\pm\pi/4}$
is not $\omega = 1$ but $\omega = \sqrt{2}$.

Figure \ref{fig:figure_2} shows the correlation functions for the
configuration of Figure \ref{fig:figure_1} in the directions of the
unit vectors $\vec e_0, \vec e_{\pi/2}$ and $\vec e_{\pi/4}$. The
first and second have been used for the reconstruction and hence show
good agreement with the reference function while the latter deviates
drastically. The correlation in direction $\vec e_{\pi/4}$ is better
described by the function
\begin{equation} \label{eq:fit}
  f(r) =
  \frac{1}{2}\left(\exp\left(-\frac{r}{8}\right)\cos\left(\sqrt{2}r\right) 
    + \exp\left(-\frac{r}{8}\right)\right) 
  .
\end{equation}
shown as the dotted line. This may be interpreted as the arithmetic
mean of the correlation function for regions (described above) with
stripes perpendicular to the direction $\vec e_{\pi/4}$ given by the
first term in Equation (\ref{eq:fit}) and the correlation function for
regions with stripes parallel to the direction $\vec e_{\pi/4}$ given
by the second term. Of course, the same correlation function is found
in direction $\vec e_{-\pi/4}$.

One step towards an isotropic reconstruction may be to use $J=4$, and
to force agreement of the correlations not only in two but in four
directions $\vec e_0$, $\vec e_{\pi/2}$, $\vec e_{\pi/4}$ and $\vec
e_{-\pi/4}$. The resulting configuration is shown in Figure
\ref{fig:figure_3}.  The pattern is significantly different from the
pattern of Figure \ref{fig:figure_1}.  There are not only stripes
parallel to the diagonal directions but more rounded formations and
concentric circles at some points. The correlation functions for
Figure \ref{fig:figure_3} are plotted in Figure \ref{fig:figure_4}.
Surprisingly we find that it is not possible to obtain agreement with
the reference correlation function.  The resulting correlation
functions in the horizontal, vertical and diagonal direction all
deviate strongly from the reference function especially at the first
minimum. Additional simulations with different cooling schedules,
damping factors and frequency of the correlation function did not give
better agreement. We have also varied the system sizes from $200
\times 200$ up to $1000\times 1000$. The results were identical.

Figure \ref{fig:sand} shows the two-dimensional reconstructions of a
Fontainebleau sandstone with porosity $\phi=0.135$. The correlation
function was truncated at $r_c=50$. The reconstruction of Figure
\ref{fig:sand}a used the correlation function only in vertical $\vec
e_0$ and horizontal $\vec e_{\pi/2}$ direction as suggested in
\cite{YT98a}. The reconstruction shown in Figure \ref{fig:sand}b,
however, calculates the full two-dimensional correlation function
according to (\ref{eq:acf_gen}) where the two-dimensional correlation
function was radially binned in the calculations to obtain a
one-dimensional function which can be compared to the one-dimensional,
isotropic correlation function of the original sandstone.  We
emphasize that in this calculation the two-dimensional correlation
function was evaluated without restrictions or simplifications.
Therefore, the calculation needed about 50 times longer than the
simplified reconstruction. Again, there are differences visible
although they are smaller than those in the reconstructions using
(\ref{eq:acf_osc}).  The shape of the pores in the full reconstruction
(Figure \ref{fig:sand}b) appear to be smoother and there is not as
much "dust" visible as in the simplified (restricted) reconstruction
shown in Figure \ref{fig:sand}a. The correlation functions plotted in
Figure \ref{fig:figure_6} reveal also that the reconstructed
micro-structure in Figure \ref{fig:sand}a is strongly anisotropic
while the micro-structure in Figure \ref{fig:sand}b is isotropic as it
should be.

In summary we note that the statistical reconstruction of porous media
with predefined two-point correlation function often requires some
reduction of numerical effort. Especially the reconstruction of
three-dimensional porous media in reasonable time does not seem
possible without such simplifications in the calculation of the
correlation function.  Of course this problem is exacerbated if one
wishes to include three-point or even higher order correlation
functions.  We applied a simplification proposed and used in
\cite{YT98a} which samples the correlation function only in certain
directions.  The effect of this simplification on the final
configurations may in some cases be negligible but in general
configurations strong anisotropy and patterns which are significantly
different from those of a proper isotropic reconstruction may appear
as a result.

\section*{Acknowledgments}
We thank B. Biswal for comments and helpful discussions, K.
H\"ofler, S. Schwarzer and M. M\"uller for technical advise and
significant parts of the C++ code, the Deutsche
Forschungsgemeinschaft for financial support through the GKKS
Stuttgart.


\begin{thebibliography}{10}

\bibitem{hil91d}
R. Hilfer, Phys. Rev. B {\bf 44},  60  (1991).

\bibitem{hil92a}
R. Hilfer, Phys. Rev. B {\bf 45},  7115  (1992).

\bibitem{hil93b}
R. Hilfer, Physica A {\bf 194},  406  (1993).

\bibitem{hil94b}
R. Hilfer {\it et~al.}, Physica A {\bf 207},  19  (1994).

\bibitem{hil95d}
R. Hilfer, Advances in Chemical Physics {\bf XCII},  299  (1996).

\bibitem{hil98a}
B. Biswal, C. Manwart, and R. Hilfer, Physica A {\bf 255},  221  (1998).

\bibitem{STJLJA94}
P. Spanne {\it et~al.}, Phys. Rev. Lett. {\bf 73},  2001  (1994).

\bibitem{YT98a}
C. Yeong and S. Torquato, Phys. Rev. E {\bf 57},  495  (1998).

\bibitem{rob97}
A. Roberts, Phys.Rev.E {\bf 56},  3203  (1997).

\bibitem{adl92}
P. Adler, {\em Porous Media} (Butterworth-Heinemann, Boston, 1992).

\bibitem{qui84}
J. Quiblier, J. Colloid Interface Sci. {\bf 98},  84  (1984).

\end{thebibliography}

\clearpage
\section*{Figure captions}
\begin{CAPTION}


\CAP{figure_1}{Simplified reconstruction of the damped
  oscillating correlation function given in Equation
  (\ref{eq:acf_osc}) by restricting the correlation function
  evaluation to the horizontal and vertical directions.}


\CAP{figure_2}{Correlation functions for the reconstruction of
  Figure \ref{fig:figure_1}. The solid line is the reference function
  of the reconstruction given in Equation (\ref{eq:acf_osc}). $+$ and
  $\times$ are the values of the correlation function of Figure
  \ref{fig:figure_1} along the horizontal and vertical direction,
  respectively. The values of the correlation function in direction
  $\vec e_{\pi/4}$ are plotted with $\circ$. The dotted line is the
  estimate given in Equation (\ref{eq:fit}) for the correlation
  function along the $\vec e_{\pi/4}$-direction. }

 
\CAP{figure_3}{Simplified reconstruction of the damped
  oscillating correlation function of Equation (\ref{eq:acf_osc}) by
  restricting the correlation function evaluation to the four $\vec
  e_0$, $\vec e_{\pi/2}$, $\vec e_{\pi/4}$ and $\vec e_{-\pi/4}$
  directions.}


\CAP{figure_4}{Correlation functions for the reconstruction of
  Figure \ref{fig:figure_3}. The reference function is plotted as
  solid line. Note the mismatch in the first minimum and maximum.}


\CAP{sand}{{\bf (a)} Two-dimensional simplified reconstruction of the
  correlation function of a Fontainebleau sandstone by restricting
  the correlation function evaluation to the horizontal and
  vertical directions.  {\bf (b)} Reconstruction with the same
  reference correlation function as in (a) but complete evaluation
  of the correlation function, i.e. without directional
  restrictions.}


\CAP{figure_6}{Correlation function for the reconstructions of
  Figure \ref{fig:sand}. The reference function is plotted as solid
  line. The correlation functions calculated in horizontal, vertical
  and diagonal direction refer to the configuration of Figure
  \ref{fig:sand}a. The complete correlation function is calculated
  from the configuration of Figure \ref{fig:sand}b.}

\end{CAPTION}



\FIGURE{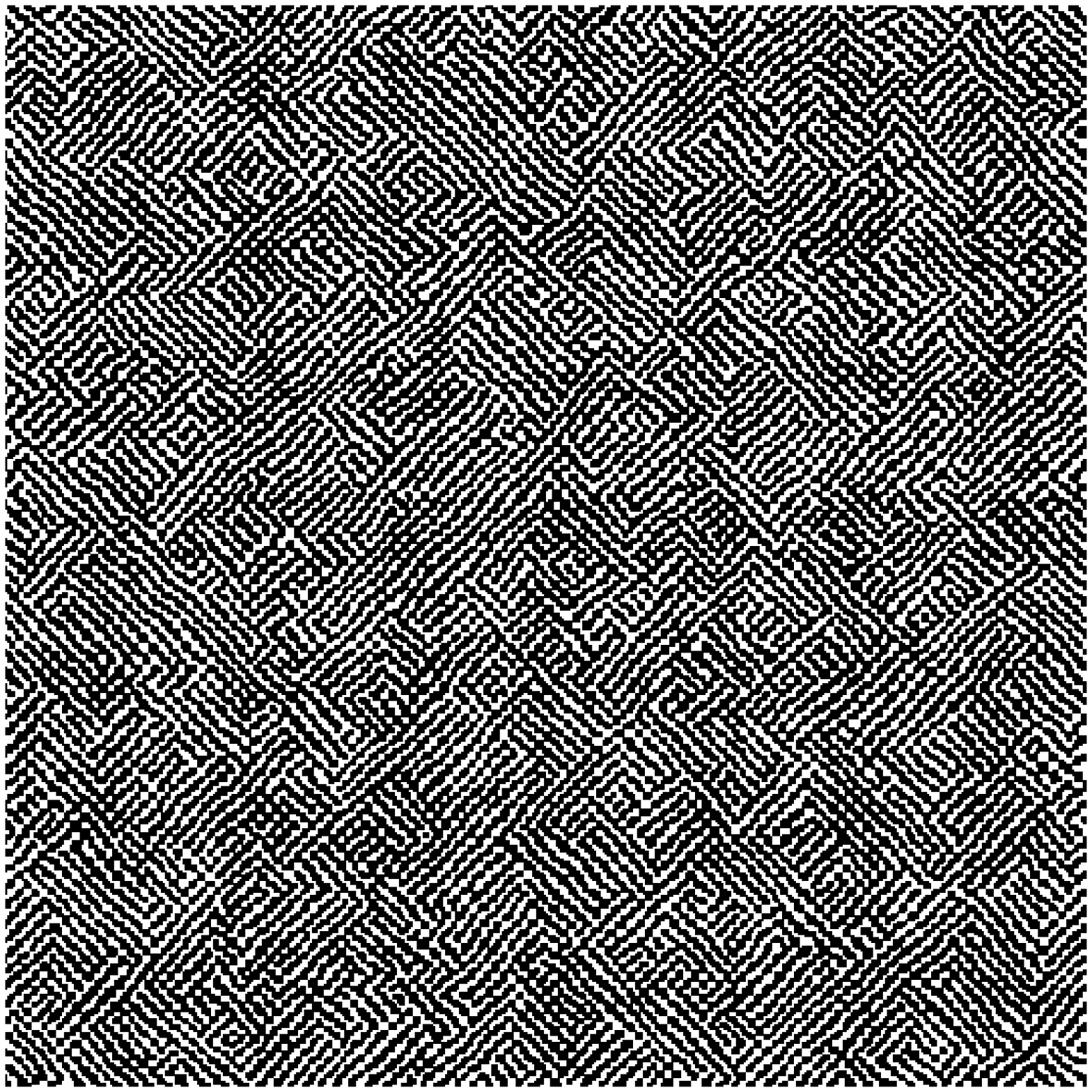}{0,65}


\psfrag{A1.x}[][b]{\Large $r$ (in units of lattice spacing)}
\psfrag{A1.y}[][]{\Large $g(r)$}
\psfrag{A1.l5}[r][r]{reference}
\psfrag{A1.l1}[r][r]{$\vec e_0$}
\psfrag{A1.l2}[r][r]{$\vec e_{\pi/2}$}
\psfrag{A1.l3}[r][r]{estimate}
\psfrag{A1.l4}[r][r]{$\vec e_{\pi/4}$}

\FIGURE{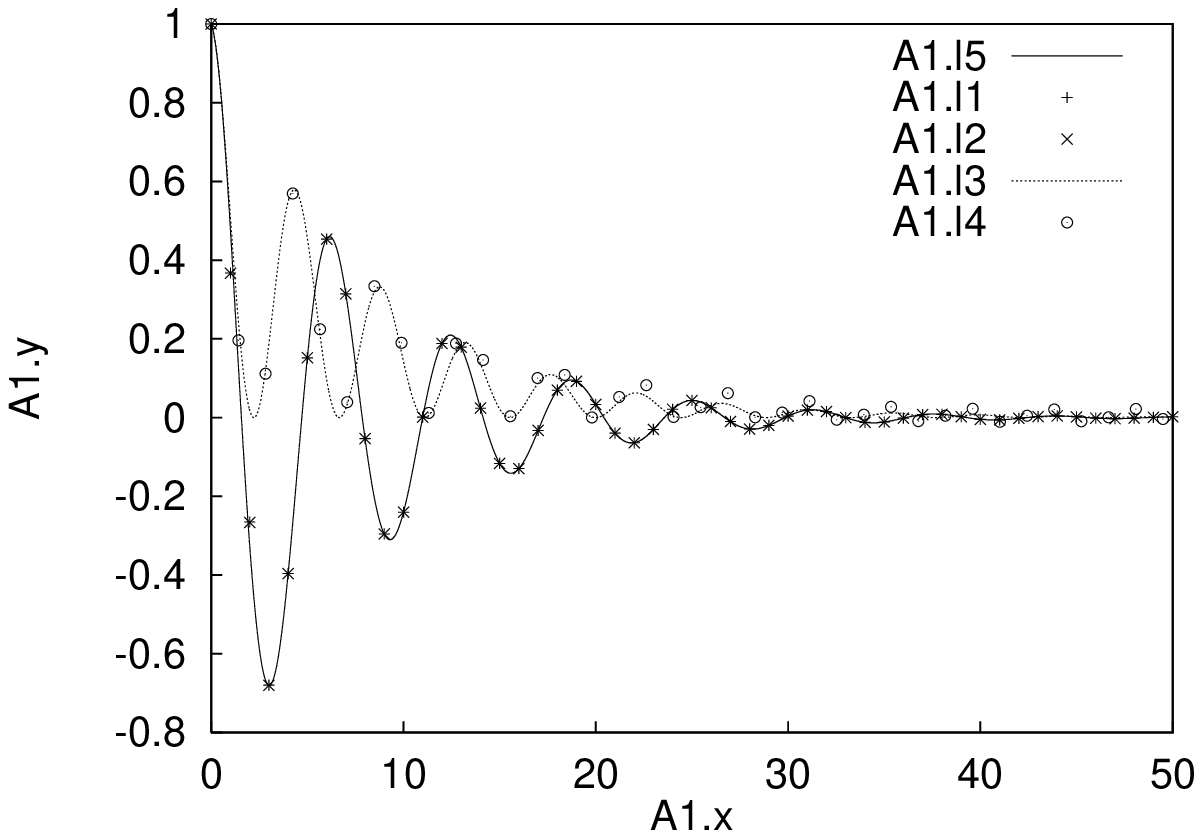}{1,0}

 
\FIGURE{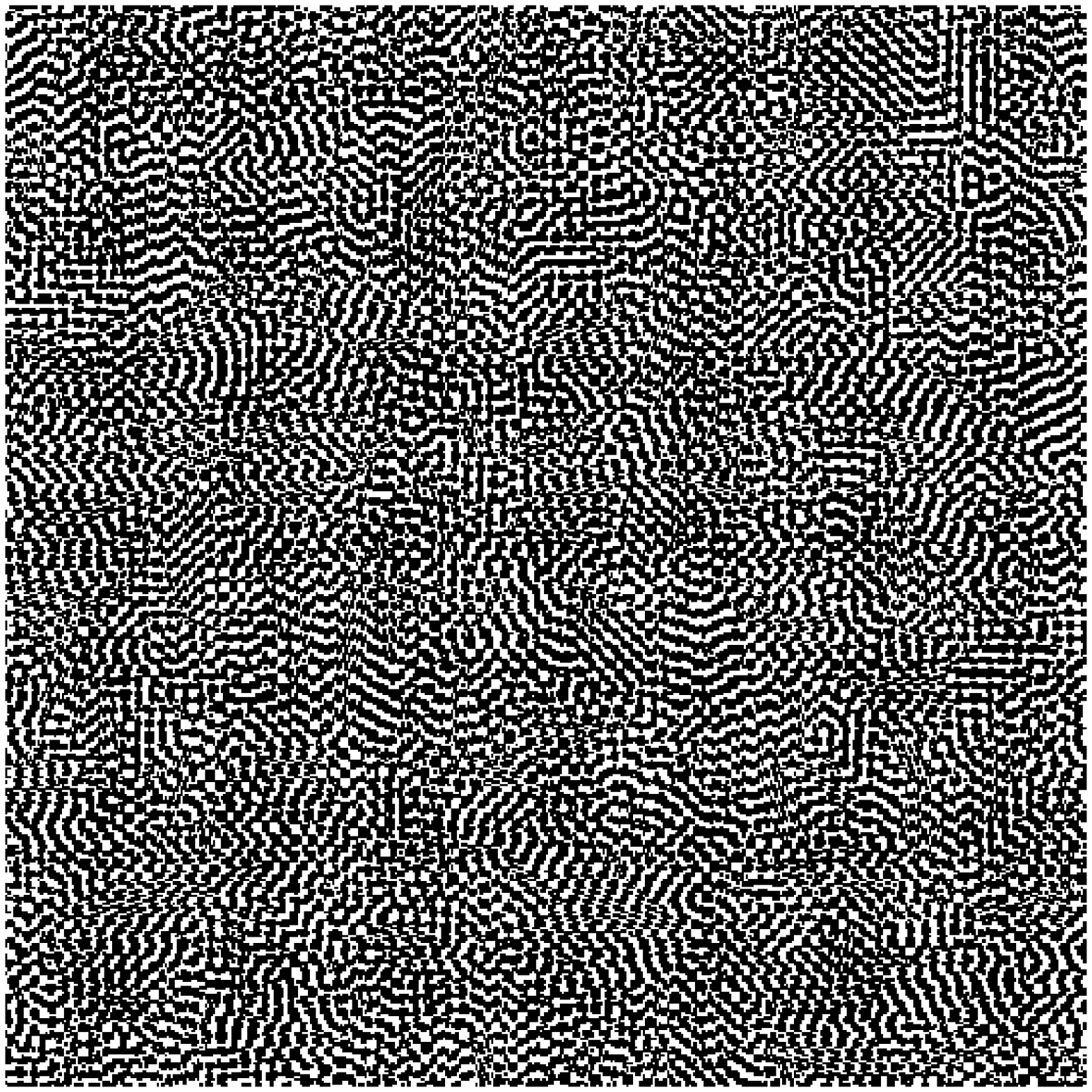}{0,65}


\psfrag{C1.x}[][b]{\Large $r$ (in units of lattice spacing)}
\psfrag{C2.y}[][]{\Large $g(r)$}
\psfrag{C1.l5}[r][r]{reference}
\psfrag{C1.l1}[r][r]{$\vec e_0$}
\psfrag{C1.l2}[r][r]{$\vec e_{\pi/2}$}
\psfrag{C1.l3}[r][r]{$\vec e_{\pi/4}$}
\psfrag{C1.l4}[r][r]{$\vec e_{-\pi/4}$}

\FIGURE{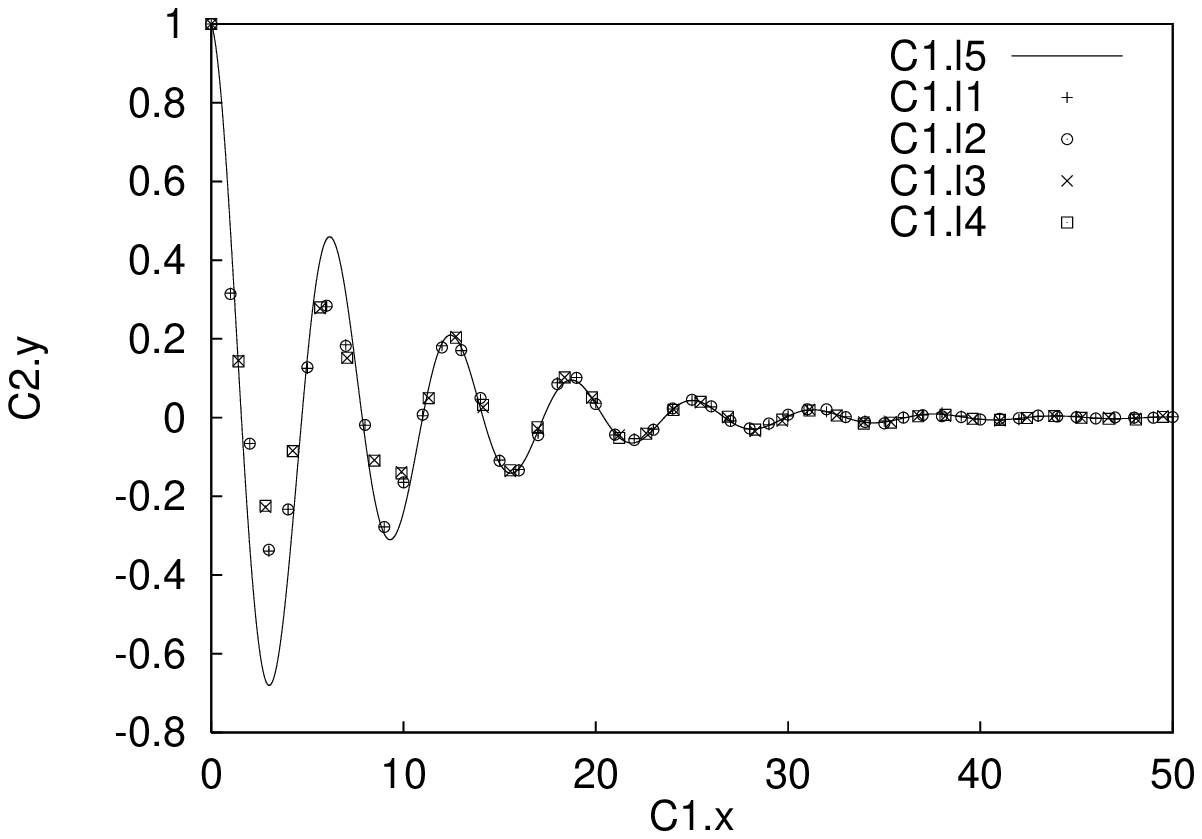}{1,0}


\clearpage

\voffset-2cm
\begin{center}
  {\small\sffamily C.Manwart and R.Hilfer \hfill Figure \ref{fig:sand}}\\[1cm] 
  \epsfig{figure=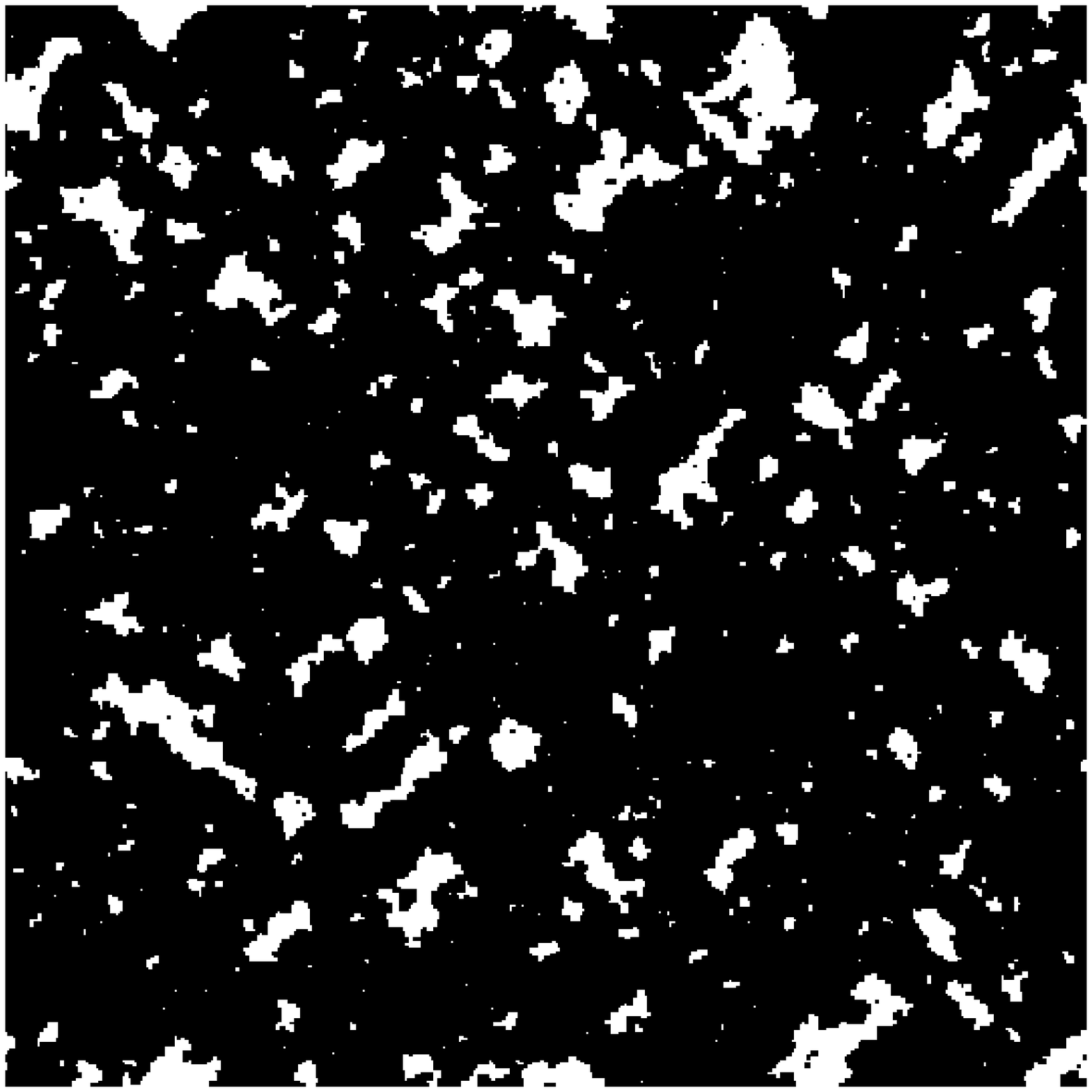,width={0.65\linewidth}}
  {\bf (a)}
  
  \epsfig{figure=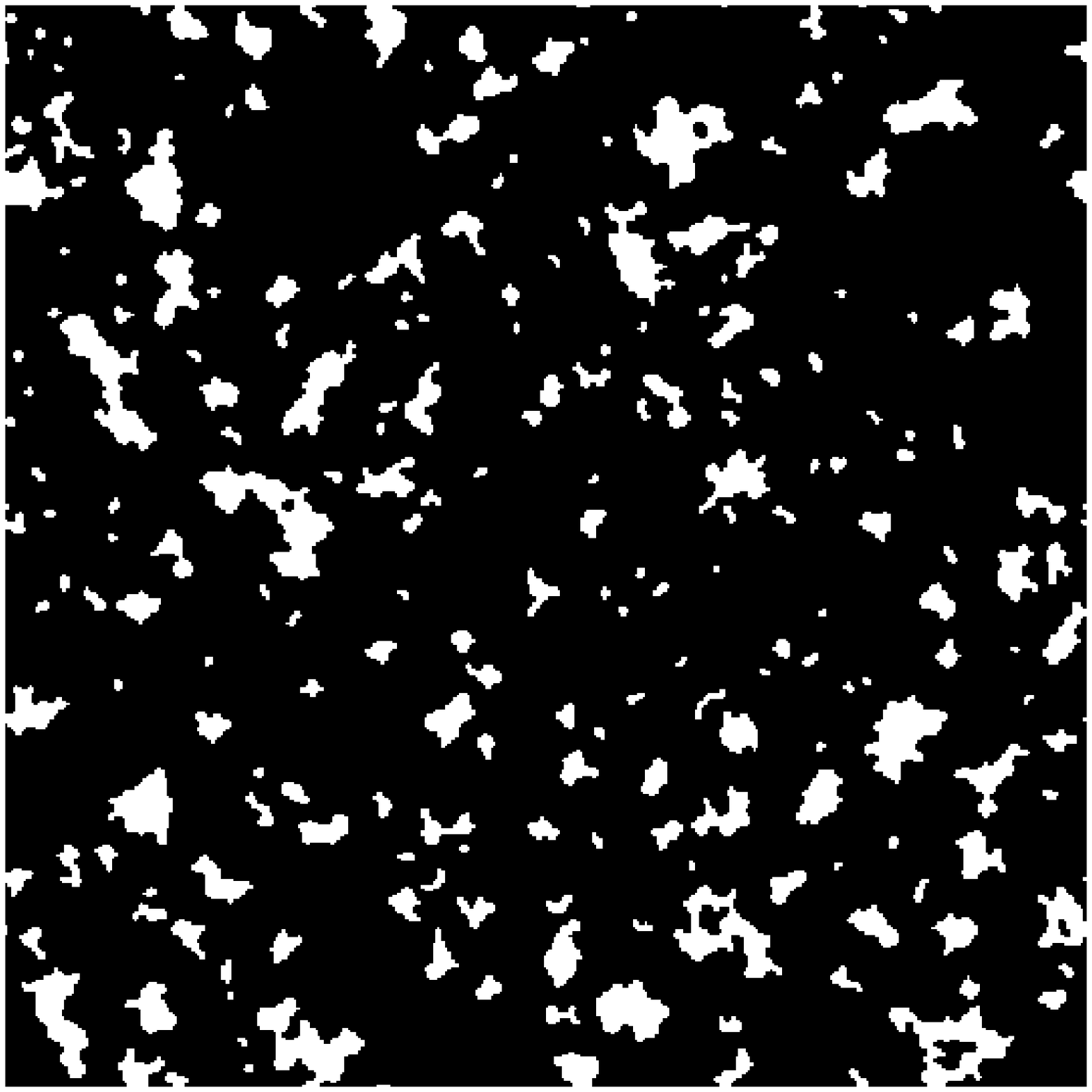,width={0.65\linewidth}}
  {\bf (b)}
\end{center}


\psfrag{E1.x}[][b]{\Large $r$ (in units of lattice spacing)}
\psfrag{E1.y}[][]{\Large $g(r)$}
\psfrag{E1.l1}[r][r]{reference}
\psfrag{E1.l2}[r][r]{$\vec e_0$}
\psfrag{E1.l3}[r][r]{$\vec e_{\pi/2}$}
\psfrag{E1.l4}[r][r]{$\vec e_{\pi/4}$}
\psfrag{E1.l5}[r][r]{complete}

\FIGURE{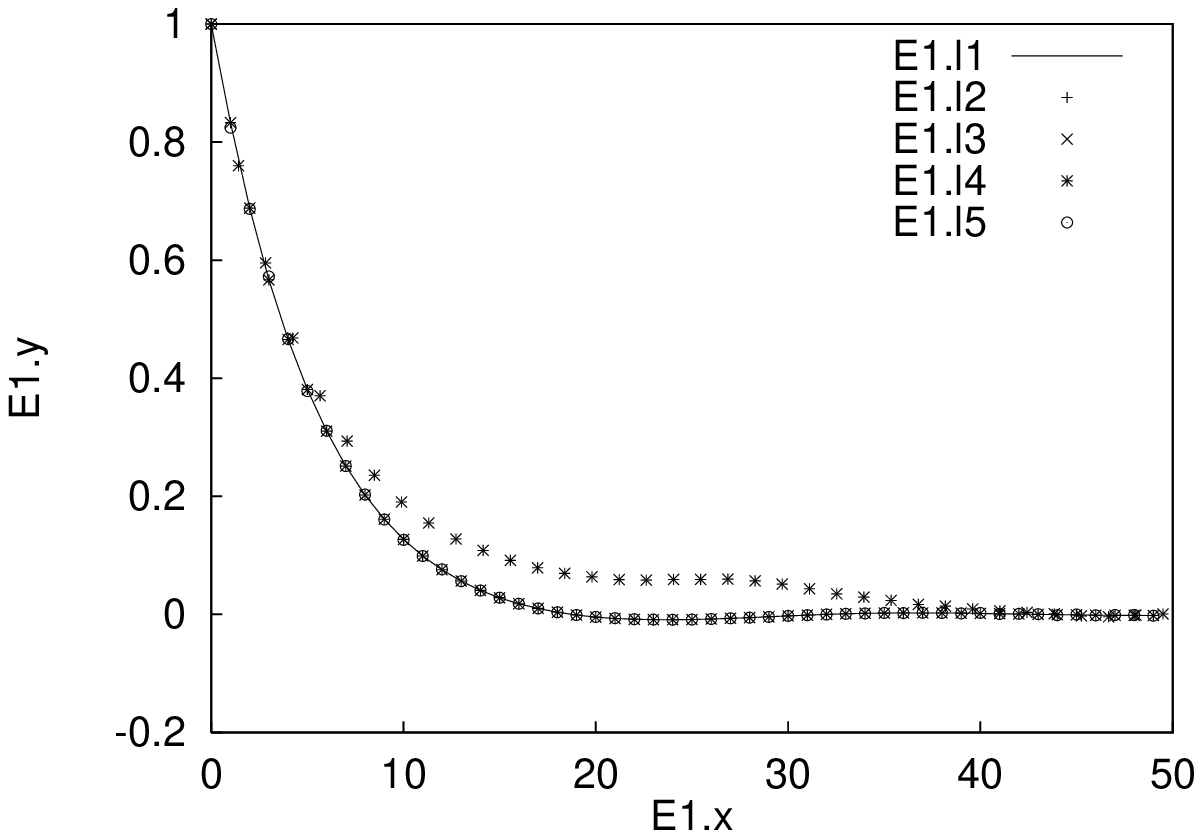}{1,0}

\end{document}